\documentclass[%
 aip,
 amsmath,amssymb,
 reprint,%
]{revtex4-1}

\usepackage{graphicx}
\usepackage{dcolumn}
\usepackage{bm}

\usepackage[utf8]{inputenc}
\usepackage[T1]{fontenc}
\usepackage{mathptmx}
\usepackage{siunitx}
\usepackage{subfig}
\usepackage{color}

\begin{document}

\preprint{AIP/123-QED}

\title{Thermal Simulation of Millimetre Wave Ablation of Geological Materials}

\author{A. Z. Zhang}
\author{S. T. Millmore}%
\email{stm31@cam.ac.uk}
\author{N. Nikiforakis}
\affiliation{ 
Cavendish Laboratory, Department of Physics, University of Cambridge, Cambridge, United Kingdom
}%


\date{\today}

\begin{abstract}
  This work is concerned with the numerical simulation of ablation of
  geological materials using a millimetre wave source.  To this end, a
  new mathematical model is developed for a thermal approach to the
  problem, allowing for large scale simulations, whilst being able to
  include the strong temperature dependence of material parameters to
  ensure accurate modelling of power input into the rock.  The model
  presented is implemented within an adaptive meshing framework, such
  that resolution can be placed where needed, for example at the
  borehole wall, to further improve the computational efficiency of
  large scale simulations.  This approach allows for both the heating
  of the rock, and the removal of evaporated material, allowing rate
  of penetration and the shape of the resulting borehole to be
  quantified.  The model is validated against experimental results,
  which indicates that the approach can accurately predict
  temperatures, and temperature gradients within the rock.  The
  validated model is then exercised to obtain initial results
  demonstrating its capabilities for simulating the millimetre wave
  drilling process.  The effects of the conditions at the surface of
  the rock are investigated, highlighting the importance of
  understanding the physical processes which occur between the wave
  guide and the rock.  Additionally, the absorptivity of the rock, and
  the impact this has on the evaporation behaviour is considered.
  Simulations are carried out both for isotropic rock, and also for a
  multi-strata configuration.  It is found that strata between similar
  rock types, such as granite and basalt, absorptive properties pose
  little problem for uniform drilling.  However, larger variations in
  material parameters are shown to have strong implications on the
  evaporation behaviour of the wellbore, and hence the resulting
  structure.
\end{abstract}

\maketitle


\section{Introduction}

The use of electromagnetic (EM) waves as a technology for drilling has
been considered since the early days of laser-driven
applications~\cite{Oglesby2014}.  In particular, this process could
allow for more efficient drilling through materials which are
problematic for conventional drill bits, such as igneous rock.
However, efficiency issues with laser sources, and the energy input
requirements for melting and vaporising rock, limited development of
EM drilling technology.  Oglesby~\cite{Oglesby2014} addressed
efficiency concerns by using a gyrotron, instead of a laser, which
produces a millimetre wave
(MMW) EM signal, with frequency in the range 30-300 GHz and for which
efficiency of commercially available units can exceed
50\%~\cite{Kasugai2008}.

Through the use of MMW drilling technology, and the advent of
commercially available high-power gyrotrons,
Oglesby~\cite{Oglesby2014} identifies several benefits over
conventional methods.  For use in wellbore drilling, using MMWs is
expected to correspond to a linear increase in drilling cost with
respect to depth, whilst for conventional methods, cost increases
exponentially with depth.  Additionally, the technology is applicable
to rocks of all hardness, and high-temperature material is not a
limiting factor in potential drill depth~\cite{osti_13999,Woskov2016}.
Related to this is the fact that there is enhanced reliability and
durability with a gyrotron source, due to the fact that it does not
come in contact with the rock itself.  It is also possible that the
process can be controlled such that the molten material forms a
vitrified liner to a wellbore during the drilling process.  Finally,
in comparison to other EM sources, MMW propagation through a dusty or
particulate medium is relatively efficient, which reduces requirements
for control of the evaporated material down a wellbore.  These
advantages suggest MMWs could be a valuable technology for deep
drilling in igneous rock environments, with applications such as
geothermal energy generation and nuclear waste entombment.  The former
of these options is of particular interest since it allows energy to
be obtained through an Enhanced Geothermal Systems (EGS) approach,
requiring the drilling of hot rock with very low natural permeability
or fluid saturation\cite{Tester2006}, which are conditions
unfavourable for conventional technology.  Successful drilling within
these conditions could allow for widespread use of geothermal
energy~\cite{McClure2014}.

In order to utilise MMWs for drilling to the depths required for
geothermal energy or other applications, understanding the evaporation
process, and how it can be controlled, is essential.  The initial cost
of a high-powered gyrotron limits the number of experimental studies
into these techniques, though Woskov and co-authors have run a series
of laboratory-scale experiments on a variety of rocks, including
basalt, granite, limestone and
sandstone~\cite{Woskov2009,Woskov2012,Woskov2013,Woskov2014,Oglesby2014,Woskov2016}.
Even for these experiments, in many cases there was insufficient power
used to achieve full vaporisation, though the capabilities of the
process to heat and melt these materials could be investigated.  Due
to these experimental challenges, and the lack of information that
will be available from down-well situations, numerical modelling
allows for further insight to understand and guide experimental work.

In this work, a thermal model of the drilling process is developed,
allowing for simulation of the MMW evaporation process over long
length and time scales. This model allows for the change in material
properties of the rock over the temperature ranges of interest,
ambient conditions to evaporation, to be considered.  Particular focus
is placed on granite and basalt since these rocks are typically
suitable for EGS technology, due to the difficulties in using
conventional drilling techniques in these
rocks~\cite{Tester2006,McClure2014}.  In addition to the
temperature-dependent thermodynamic properties of these rocks, the
dependence on MMW absorption is also considered, which effectively
transitions from volumetric to surface heating of the material with
increasing temperature.

The rest of this paper is laid out as follows;
section~\ref{sec:methodology} describes the mathematical and physical
formulation of the model, including material parameters of the rocks
considered, and the numerical methods for simulating the MMW drilling
process are given in section~\ref{sec:numerical-approach}.
Section~\ref{sec:validation} provides a validation study,
demonstrating the ability of the numerical model to reproduce
experimental measurements.  Section~\ref{sec:results} considers
further evaluation of the approach for a single material,
investigating changes in properties of the rock, whilst
section~\ref{sec:multi-strata-modell} introduces multiple strata,
considering the effects of two layers of material, with an angled
interface between them.  Finally, conclusions and further work are
described in section~\ref{sec:conclusions}.

\section{Mathematical formulation}
\label{sec:methodology}

The overall problem of rock ablation through MMW drilling is complex,
and involves multiple time and length scales.  The full process
involves phase change from solid to molten rock, and subsequently to
vapour, at which point gas flow (typically an inert gas) carries
vaporised material away from the bottom of the wellbore.
Additionally, molten rock will flow under the induced thermal
gradients and reflection and scattering of the MMW source makes
modelling energy input into the rock a complex process.  Finally, the
physical and absorptive properties of rock are strongly, and
non-linearly, temperature dependent.

In order to simulate this process at large scales, such that rate of
penetration can be measured, potentially thorough multiple material
strata, this work assumes that the overall drilling process is
governed by a thermal model.  Gas flow occurs on a time scale too
short for this scale, and can be considered a boundary condition,
whilst material flow is assumed to occur over length scales too small
to allow for efficient large scale simulations.  This reduces the
complexity of the system of equations required to describe the MMW
drilling process, however the material properties cannot be simplified
in such a manner.  The non-linear thermal treatment of the rock is
still included in the model presented here.  Scattering of the MMW
source is not directly considered in this work, but rather, a
simplified energy input is employed.  However, since energy input
occurs through a source term, the model presented here is compatible
with more advanced techniques for determining the MMW source
behaviour.
\\

The aim of this model is to enable the simulation of the evaporation
process over long time scales (10s of seconds upwards) and large
length scales.  In order to achieve this with reasonable computational
efficiency, it is assumed that material transport effects are
negligible, and pressure is constant.  Under these assumptions, the
system reduces to a single thermal equation,

\begin{equation}
  \label{eq:heat-eq}
  \rho(T) c_p(T) \frac{\partial T}{\partial t} = \nabla \cdot \left(
    \kappa(T) \nabla T\right) + \dot{q},
\end{equation}

where, $T$ is temperature, $\rho$ density, $c_p$ specific heat at
constant pressure, $\kappa$ thermal conductivity and $\dot{q}$ is a
heat transfer term.

This final term comprises the heat input from the
MMW source, as well as contributions from radiation and convection,
and is given by
\begin{equation}
  \label{eq:q-dot-tot}
  \dot{q} = \dot{Q} - \varepsilon \sigma \left(T^4 - T^4_0\right) - h\left(T-T_0\right)
\end{equation}
where $\dot{Q}$ is the input from the MMW source, $T_0$ is the
temperature above the surface, $\varepsilon$ is the emissivity of the
surface, $\sigma$ is the Stefan-Boltzmann constant and $h$ is the air
convection coefficient.

The input, $\dot{Q}$, is given by
\begin{equation}
  \label{eq:heat-input-general}
  \dot{Q} = P({\bf x}) \alpha_\lambda(T) \exp \left(-\alpha_\lambda(T) (z_0-z)\right)
\end{equation}
where $P({\bf x})$ is the power incident on the surface, $z_0 =
z_0(x,y,t)$ is the height of the surface, and
$\alpha_\lambda (T)$ is the absorption coefficient of the medium,
which may itself depend on temperature.  The incident power is
dependent on both the MMW profile, which, in practice, emerges from a
wave guide above the material surface, and also on the distance to this
surface.  In this work, it is assumed that the source profile is
Gaussian, and that it is aligned with the $z$-axis, giving an incident
surface power of
\begin{equation}
  \label{eq:inc-sur-pow}
  P = I_0 \left(\frac{\omega_0}{\omega(z_0-z)}\right)^2\exp\left(\frac{-2 r^2}{\omega(z_0-z)^2}\right)
\end{equation}
where $I_0$ is the peak intensity, $\omega_0$ is the wave guide radius,
$\omega(z_0-z)$ is the radius of the beam at which intensity is given by
$I = I_0 e^{-2}$ and, for a wave guide centred on $(x_0,y_0)$, $r^2 =
(x-x_0)^2 + (y-y_0)^2$.  The peak intensity is given by
\begin{equation}
  \label{eq:peak-intens}
  I_0 = \frac{2P_0}{\pi \omega^2}
\end{equation}
where $P_0$ is the peak incident power, and $\omega(z_0-z)$ is computed
through
\begin{equation}
  \label{eq:omega-z}
  \omega(z_0-z) = \omega_0 \sqrt{1 + \left(\frac{z_0-z}{z_R}\right)^2}
\end{equation}
where $z_R$ is the Rayleigh range, which is related to the source
wavelength, $\lambda$, through
\begin{equation}
  \label{eq:ray-range}
  z_R = \frac{\pi \omega^2_0 n}{\lambda}
\end{equation}
and in this work, $n=1$ is used.  This approach, though relatively
simple, does allow the stand-off distance between the surface and the
wave guide to be considered, allowing some dispersion, and absorption
from the gas, over this distance.

Here, it is noted that experimental tests have shown that the MMW
source is capable of breaking down the air between wave guide and
surface, generating a plasma~\cite{Oglesby2014,Woskov2016}.  This is
substantially more absorptive than unionised air, and can reduce the
incident power on the surface by three orders of magnitude.  However,
in practice, inert gas flow will be used to prevent plasma formation,
hence these effects are not incorporated within the model presented
here; it is assumed that all absorption from this gas is based on
equilibrium conditions.

The thermal model, given by equation~(\ref{eq:heat-eq}), is augmented
with a material removal model, dealing with the case in which the
evaporation temperature of rock is exceeded.  This technique assumes
that once material has evaporated, it is entrained within gas flow
at the surface, and thus will not condense again at, or close to,  the
evaporation front.
Within the overall model, this behaviour is achieved by altering the
properties of any region of the computational domain which reaches, or
exceeds, the evaporation temperature of the rock.  Once this condition
has been reached, the material properties are then altered to be that
of the surrounding gas.  To implement this process, and additional
scalar variable is used to describe the material in each cell, e.g. a
rock (of which there may be multiple types) or gas.
\\

Simulation of MMW drilling requires properties for rock to be
understood from ambient conditions up to their vaporisation
temperature (in excess of 3,000 K).  Over these ranges, density,
thermal conductivity, specific heat and absorptivity all vary
substantially.  For lower temperature ranges, up to around 1,300 K,
these properties are well studied, see, for example,
Heuze~\cite{Heuze1983}, Hartlieb et al.~\cite{Hartlieb2016},
Branscome~\cite{Branscome2006} and Waples and
Waples~\cite{Waples2004}.  However, above these ranges, properties are
not well known, in part due to the high experimental cost for such
investigations.  For many thermal properties, Oglesby has found that
using simple linear or constant extrapolations from known behaviour is
typically sufficient to give reasonable high-temperature
behaviour~\cite{Oglesby2014}.  Such behaviour can be augmented where
necessary through studies on magma, undertaken by the United States
Geological Serviced, detailed by Robertson~\cite{Robertson1988}.

Density of geological materials as a function of temperature may be
known for temperatures below $\sim 1000$ K but above this, few
measurements exist.  In this work, density is given by~\cite{Oglesby2014}
\begin{equation}
  \label{eq:dens-eq}
  \rho = a_\rho (T - 273.15) + b_\rho
\end{equation}
where $a_\rho$ and $b_\rho$ are constants.  This equation holds up
until a cut-off value (determined by the limits of experimental
measurements), after which a constant extrapolation is used, to ensure
density values do not drop unphysically low.

Thermal conductivity is treated in a similar manner, in this case,
however, two possible functional forms exist~\cite{Oglesby2014};
\begin{equation}
  \label{eq:th-cond-rec}
  \kappa = \frac{1}{a_\kappa + b_\kappa (T - 273.15)}
  \end{equation}
  or
\begin{equation}
  \label{eq:th-cond-eq}
  \kappa = c_\kappa (T - 273.15)^2 + d_\kappa (T - 273.15) + e_\kappa
\end{equation}
with constant parameters, $a_\kappa$, $b_\kappa$, $c_\kappa$,
$d_\kappa$ and $e_\kappa$.  As with density, above a threshold,
thermal conductivity is treated as constant.

Specific heat is treated slightly different from the other parameters,
in that it contains a contribution from experimental measurements, and
an additional contribution over phase change ranges to deal with
latent heating effects.  For the materials considered in this work,
approximations to experimental measurements use a quadratic expression
up to a threshold value, and subsequently a linear
fit~\cite{Oglesby2014}.  This gives a general form
\begin{equation}
  \label{eq:sp-heat}
  c_p =
  \begin{cases}
    a_c (T - 273.15)^2 + b_c (T - 273.15) + c_c & T \le T_c \\
    d_c (T - 273.15) + e_c & T > T_c
  \end{cases}
\end{equation}

In order to incorporate latent heat behaviour into the model, the
approach of Yang et al.~\cite{Yang2010} and Oglesby~\cite{Oglesby2014}
is followed.  Here, the latent heat of fusion, $\xi_f$, is linearly
distributed between solidus and liquidus temperatures ($T_s$ and $T_l$
respectively), and a similar treatment is used for the latent heat of
evaporation, $\xi_v$.  In this latter case, the temperature range over
which evaporation occurs is not known for rocks, and an approximation,
to distribute the latent heat effects over a 270 K temperature range,
starting from evaporation temperature, is used.  The overall specific
heat is then the sum of these two contributions.

The absorptivity of a material is dependent on the permittivity,
$\epsilon$, permeability, $\mu$, and electrical conductivity,
$\gamma$, of the material itself, and also the frequency of the
source, $\omega$.  In practice, this can be computed through
\begin{equation}
  \alpha_{\lambda} = 2  \omega\sqrt{\frac{\epsilon \mu}{2}}\left[
    \sqrt{1+\left(\frac{\gamma}{\epsilon \omega}\right)^2}-1\right]^{\frac{1}{2}}
  \label{eq:absorptionCoefficientmethod}
\end{equation}
however, the material properties for rocks are not necessarily known.
As a result, the absorptivity is typically experimentally measured,
and assumed to be piece-wise constant, potentially jumping between
solid and liquid phases~\cite{Oglesby2014,Woskov2012}.

\subsection{Treatment of quartz $\alpha$-$\beta$ transition}
\label{sec:treatm-quartz-alpha}

\begin{figure} \centering
  \includegraphics[width=0.49\textwidth]{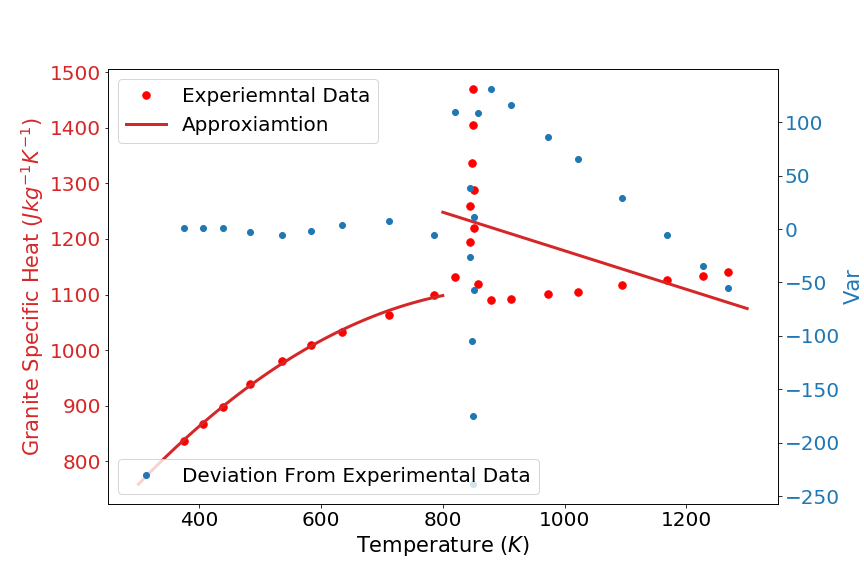}
\includegraphics[width=0.49\textwidth]{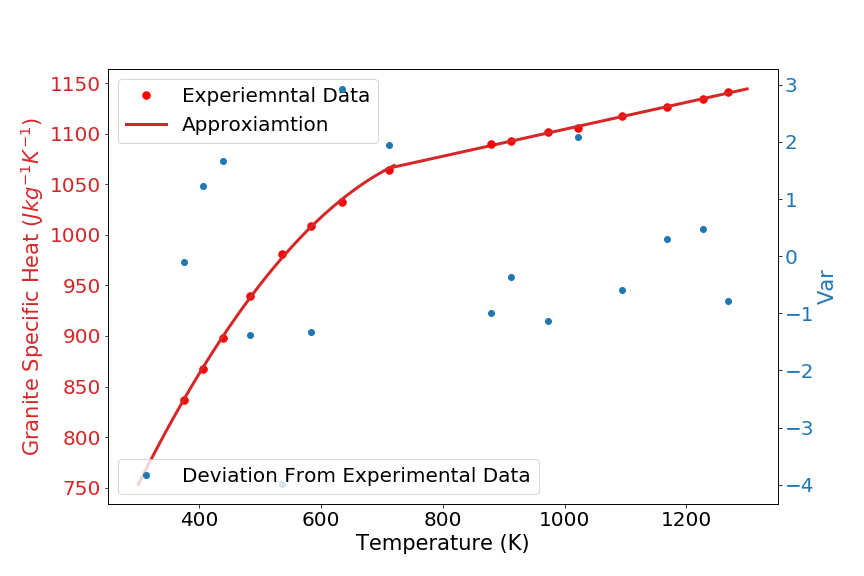}
    \caption[ Linear approximation of specific heart with and without
    transition. ]{Comparison of linear approximations to specific heat
      data both with (top) and without (bottom) taking the $\alpha$-$\beta$ phase
      transition of quartz into account.  Red points show the
      experimental data, red lines show the
      approximation to this data, and the relative error
      to measured points is shown by the blue dots.  By removing the
      phase transition behaviour from the approximation, it is clear
      that accuracy increases, thought this requires an additional
      treatment to get the correct specific heat within this
      transition region.
    }
    \label{fig:PhaseTransution }
\end{figure}

Quartz content of a geological material can have an effect on its
thermal properties since it
exhibits an $ \alpha$-$\beta$ transition at around
845~K~\cite{LeChatelier1889,Robertson1988,Dolino,Hartlieb2016}, and
this subsequently alters the specific heat of the quartz, and hence
the underlying rock~\cite{Navrotsky}.  In this work, we consider
basalt, which has a low quartz content, and thus these effects are
negligible, and granite, which has a much higher quartz content.
Here, care has to be taken, both when attempting to obtain functional
fits to experimental data for specific heat of a granite when
temperature crosses this transition region, and also in incorporating
these effects within the numerical model.
Figure~\ref{fig:PhaseTransution } compares two functional fits to
experimental data for granite, one in which the values measured over
the phase transition have not been accounted for, and a second showing
the phase transition removed.  It is clear that by removing the phase
transition effects from this functional fit, the overall error in
approximating the specific heat can be reduced.  However, in order to
model the behaviour within this region, specific heat is then
increased within the transmission region, using a similar technique as
is used to incorporate latent heat effects.

\section{Numerical approach}
\label{sec:numerical-approach}

The numerical techniques used to solve the model formulated in
section~\ref{sec:methodology} have been designed to deal with the
length and time scales of the problems of interest.  The absorptivity
of materials such as basalt and granite, described by
equation~(\ref{eq:heat-input-general}), are such that a significant
portion of the incident energy is absorbed over a short length scale
when compared to the diameter of the incident beam.  As a result,
there is a strong temperature gradient at the evaporation front, which
requires suitable computational resolution to be modelled accurately.
To achieve this desired resolution, whilst retaining computational
efficiency, a finite volume formulation is used for solving
equation~(\ref{eq:heat-eq}), upon which hierarchical adaptive mesh
refinement (refining in both space and time) can be implemented.  In
addition to solving for the temperature profile, two additional
variables are updated; the material identifier required for the
material removal model, and the vertical depth beneath the surface of
each position within the computational domain.  This latter variable
is used in computing the MMW beam absorption in
equation~(\ref{eq:heat-input-general}).

In order to solve equation~(\ref{eq:heat-eq}), and to update these
additional properties, a three-step process is implemented:
\begin{enumerate}
\item Using the data at the current time $t = t^n$, compute the power incident
  upon the current surface, and use this to update the temperature to 
  time $t^{n+1} = t^n + \Delta t$ for a given time step $\Delta t$.
  To do this, equation (\ref{eq:heat-eq}) is cast into canonical form
  \begin{equation}
    \label{eq:amrex-canonical}
    \left( A \alpha + B \nabla \cdot \beta \nabla\right) T = f
  \end{equation}
  where $A$ and $B$ are scalars, and $\alpha$, $\beta$ and $f$ are
  scalar fields.  The solution is obtained using the biconjugate
  gradient stabilised method (BiCGSTAB)~\cite{van1992bi}.
\item For any volumes in the computational domain with $T > T_v$, set
  the properties in this cell to be those of the surrounding gas.  It
  is assumed that there is a gas flow over the surface of the material
  which removes vaporised and particulate matter, for which the
  surface treatment described in section~\ref{sec:surface-heat-losses}
  is used.
\item Use the interface between the geological material and the
  surrounding gas to recompute the depth beneath the surface, using the
  techniques described in section~\ref{sec:determ-surf-height}. 
\end{enumerate}

\subsection{Determining surface height}
\label{sec:determ-surf-height}

In order to correctly prescribe the volumetric heating term of
equation~(\ref{eq:heat-input-general}), the depth beneath surface is
required.  The surface, $z_0$, evolves with time through material
evaporation, and thus an additional technique is required to track the
depth.  In this work, a signed distance function, $\phi$, is used,
which satisfies
\begin{equation}
  \label{eq:sdf}
  \left|\frac{\partial \phi}{\partial z}\right| = 1
\end{equation}
and the zero contour, $\phi({\bf x}) = 0$, describes the current surface
of the rock.

In order to compute the signed distance function, a fast sweeping
algorithm is used~\cite{Zhao2004}.  This requires knowledge of the
current location of the interface; within this work, the it is known
whether a computational cell is either rock (molten or solid) or
surrounding gas, hence the interface can be set to be at the cell
boundary between the transition from rock to gas.

\subsection{Surface heat losses}
\label{sec:surface-heat-losses}

One of the key challenges in ensuring accurate melt treatments, and
for validating the methods presented in this work, is the treatment of
the surface conditions, which are not well studied for geological
materials under temperatures such as those achieved through a MMW
source.  The convective coefficient, $h$, depends not only on the
temperature of the underlying material, but also on the conditions at
the surface, and will vary depending on the level of gas flow, with a
potential range of $h \in [25,250]$ W/m$^2$ K.  Similarly, the surface
losses are affected by the temperature of the material above the
surface ($T_0$ in equation~(\ref{eq:q-dot-tot})).  Even with air flow
over the surface, the high temperatures beneath the MMW source will
cause local temperature to exceed ambient conditions.

In order to ascertain suitable conditions for surface heat losses,
several cases were computationally considered with a granite
substrate, varying the convection coefficient (both for solid and
molten material), as well as $T_0$.  It is noted that it is not
computationally efficient to solve the heat loss problem at the
surface; this requires considering the gas flow, and occurs at time
scales substantially shorter than those of material evaporation, and
would hence have a detrimental effect on computational efficiency.  It
was found that assuming a surface temperature of
\begin{equation}
  \label{eq:surface-temp}
  T_0 = \mathrm{min}\left(T_{\mathrm{initial}}, 0.86 T_{\mathrm{surface}}\right)
\end{equation}
was a straightforward, yet accurate, technique to deal with these
effects, where $T_{\mathrm{surface}}$ is the temperature of the
material at the surface.  It is noted that this approach did not
consider the possible increase in the convective heat transfer
coefficient at higher temperatures, though improving this aspect of
the model would require further experimental study.

\subsection{Adaptive mesh refinement}
\label{sec:adapt-mesh-refin}

\begin{figure}
  \centering
  \includegraphics[angle=0,width=0.7\linewidth]{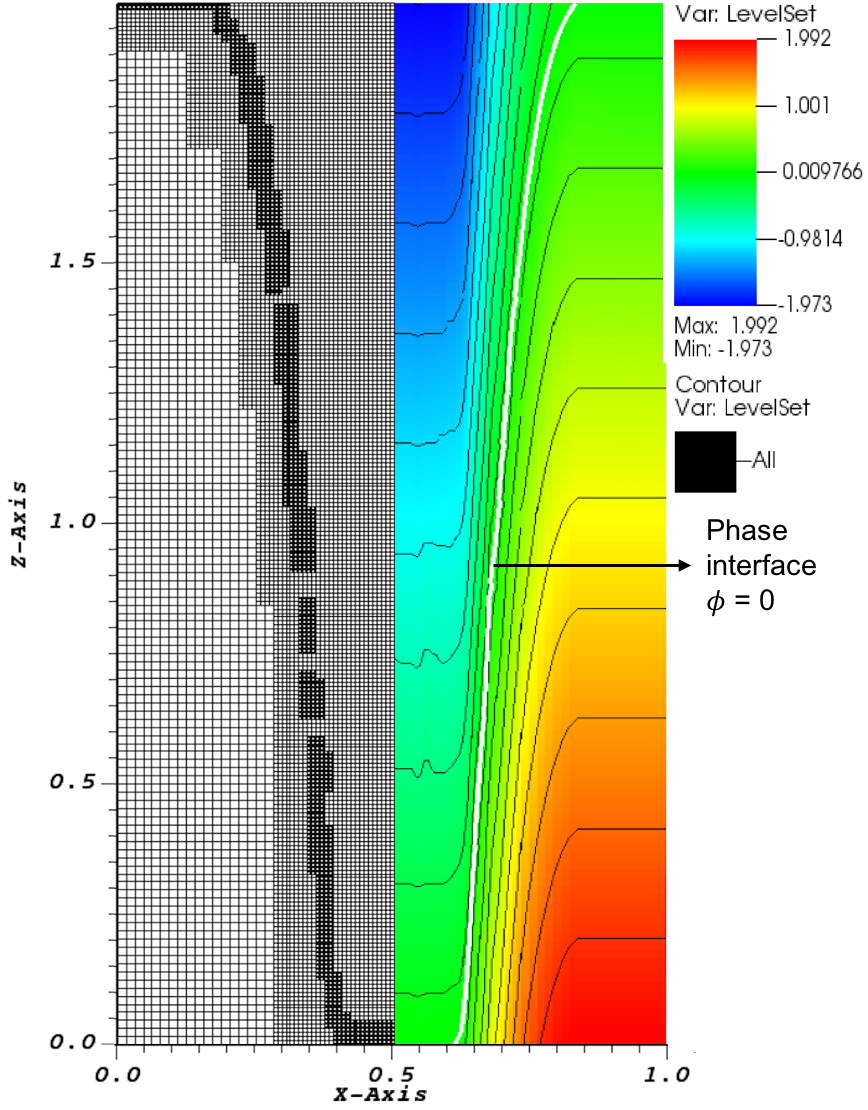}
  \caption{Demonstration of the application of AMR and of the level
    set function for recording material depth.  AMR is applied based
    on temperature and material interface parameters, allowing the
    regions of strongest temperature gradient to be captured at the
    highest resolution.  It is noted that the contours within the
    wellbore do suffer from visualisation artefacts caused by the
    automatic interpolation of the contour location.}
    \label{fig:EM2}
\end{figure}

The numerical methodology described in this section is inherently
suitable for hierarchical adaptive mesh refinement (AMR)~\cite{Bell94}.
This technique allows for regions of the domain to be both refined in
space and in time, hence regions of the computational domain far from
the energy source can be evolved with a larger time step than those
refined regions with high temperatures and strong temperature
gradients.  In this work, two thresholds are used to govern where AMR
places the additional refinement, based on the governing processes
which lead to material removal; temperature thresholds, e.g.\
$T > T_m$, and the location of the material interface.  It is noted
that these two regions will typically coincide for the ablation of
geological materials, but by refining the material interface, it is
ensured that this region is represented accurately, even when it is no
longer being directly heated, and thus allowing for accurate
quantification of the wellbore shape.

Figure~\ref{fig:EM2} shows a typical mesh
distribution, 
with the level set function also plotted to demonstrate how the
additional resolution follows the material interface.  The efficiency
of a simulation using AMR depends heavily on the proportion of the
domain covered by the refined mesh, but even in cases with a wellbore
occupying a large portion of the computational domain, simulations are
demonstrably more efficient with mesh refinement.  Sample results,
which show the improvements in computational run time, are shown in
table~\ref{tab:amr-times}.  Comparisons are made between simulations
with the same finest resolution ($32 \times 32 \times 64$ or
$64 \times 64 \times 128$ on a single CPU), with and without AMR.  The
benefits of mesh refinement are clear; using two levels of mesh
refinement, using two levels of AMR reduces run times by about 50\%.
In these cases, greater gains are not seen due to the proportion of
the domain occupied by the wellbore surface, which is always refined.
Moving to larger geometries, however, greater gains could be made,
where the wellbore only needs refinement where there is power input,
and thus these results can be thought of as the minimum gains
available from AMR.

\begin{table}[!ht]
  \centering
  \begin{tabular}{|c | c | c | c |}
    \hline
    Resolution & Simulation &  Resolution & Simulation \\
    (cells) &  time (s) &  (cells) & time (s) \\
    \hline
    $64 \times 64 \times 128$ & 9423 & $32 \times 32 \times 64$ &  1311 \\
    \hline
     $32 \times 32 \times 64$ &  &$16 \times 16 \times 32$ &\\
     1 AMR level & 7509  & 1 AMR level & 854 \\
    \hline
    $16 \times 16 \times 32$ &   &$8 \times 8 \times 16$ & \\
    2 AMR levels &  5762  & 2 AMR levels &599 \\
    \hline
  \end{tabular}
  \caption{Comparison of simulation times with and without AMR, for
    the same finest resolution.  Initial data for these tests is as
    given in table~\ref{tab:test-1-init-data}.}
  \label{tab:amr-times}
\end{table}

The implementation of AMR requires the automatic creation, or removal,
of regions of refinement, as well as consistency in the numerical
solution techniques across refinement level boundaries.  Within this
work, the formulation and numerical techniques are implemented with
the support of the AMReX framework, a mesh generation package
supporting hierarchical adaptive mesh refinement (AMR), compatible
with for elliptic, parabolic and hyperbolic
systems~\cite{Bell94,Almgren1998,Zhang2019}.

\section{Validation}
\label{sec:validation}

\begin{figure}
    \centering
    \includegraphics[width=0.49\textwidth]{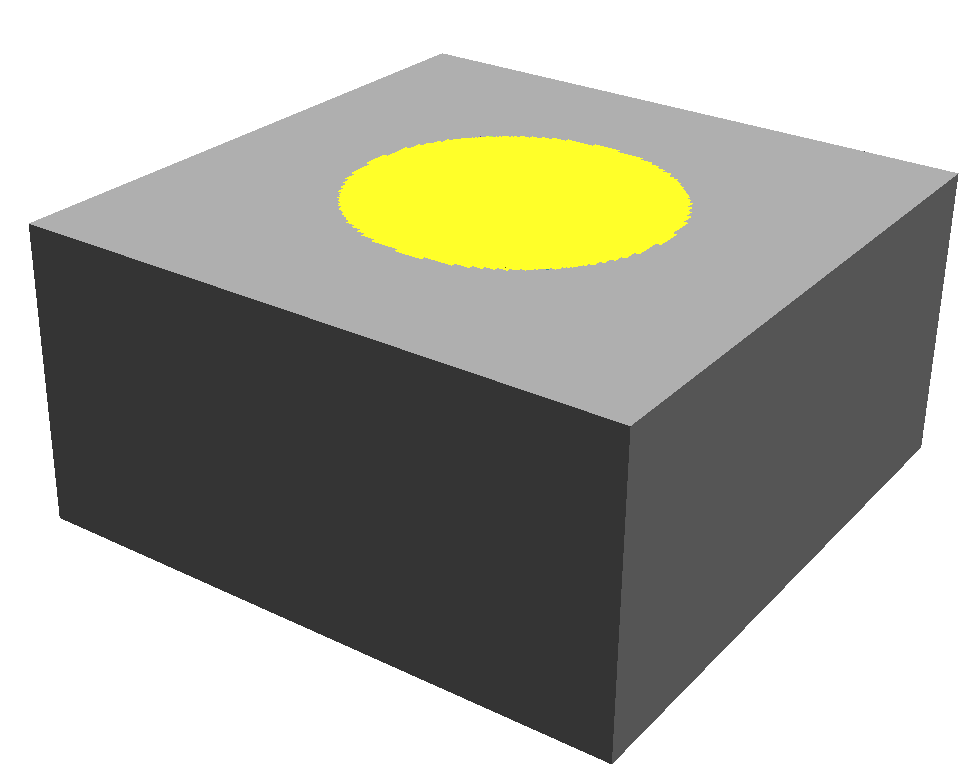}
    \caption{Representative initial configuration for the numerical
      results presented within this section.  The yellow circle shows
      the location beneath the wave guide, i.e.\ the region of incident
      MMWs.}
        \label{fig:setup}
\end{figure}

In order to validate the model described in
section~\ref{sec:methodology}, a comparison to experimental data is
considered.  Although MMW is still an emerging technology and
experimental data is scarce, the study of Oglesby~\cite{Oglesby2014}
provides detailed temperature measurements for the heating of granite
through a varying MMW source.  This experiment considers a source upon
a flat surface of granite, where the edges of the rock are
sufficiently far from the incident beam that they are not heated.  A
sample geometry used in the numerical simulations is shown in
figure~\ref{fig:setup}, where a cuboidal geometry is used for
simplicity given the underlying Cartesian coordinate system, and the
region subject to incident MMWs is shown in yellow.

\begin{figure}
    \centering
    \includegraphics[width=0.49\textwidth]{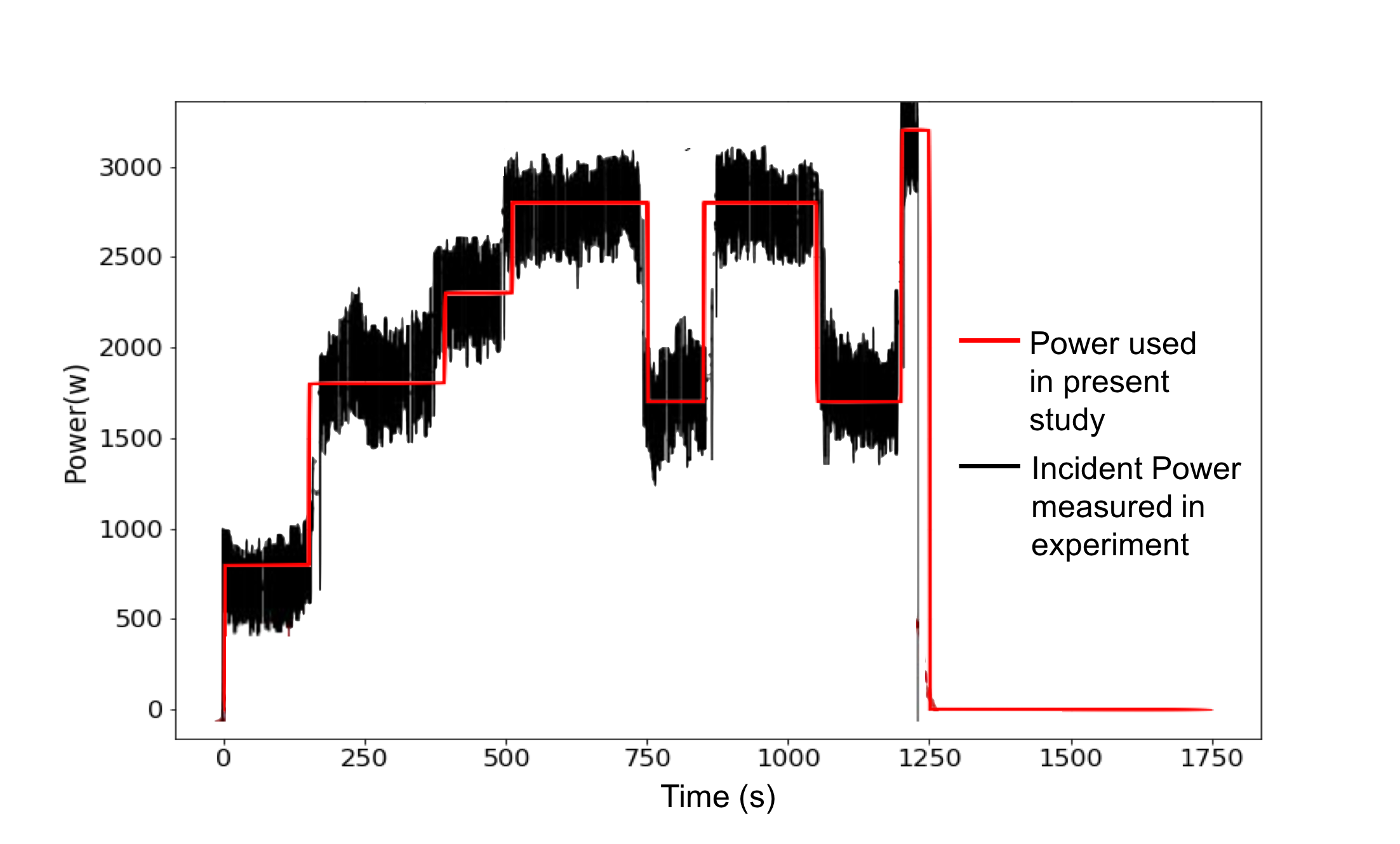}
    \caption{Comparison of the power recorded at the surface in the
      experiment of Oglesby~\cite{Oglesby2014} (black line) and the
      values used in this work (red line).  Several distinct changes
      in power are clear; the input power for this work uses an
      average through each of these regions.}
        \label{fig:Powersource}
\end{figure}

The variations in the MMW source were measured at the surface over the
course of the experiment (a duration of 30 minutes), with several
clear, sharp changes in the power of the gyrotron.  This profile is
shown in figure~\ref{fig:Powersource}, and is compared to the
approximation to this profile used in this work.  Due to the noisy
nature of the measurements, average values are taken over each
distinct region within the profile.  These values are quantified in
table~\ref{tab:validationTest_Parameters}.  The remaining parameters
for the power input and the domain used for this study are given in
table~\ref{tab:validation-test-init-data}.
\begin{table}
    \centering
    \begin{tabular}{c|c r}
    
     Time (s) \quad & \quad Power (W)  \\
    \hline
     0 - 150    \quad  & \quad  800  \\
     150 - 390  \quad  & \quad  1800 \\
     390 - 510  \quad  & \quad  2300 \\
     510 - 750  \quad  & \quad  2800 \\
     750 - 850  \quad  & \quad  1700 \\ 
     850 - 1050 \quad  & \quad  2800 \\
     1050 - 1200\quad  & \quad  1700 \\
    1200 - 1250 \quad  & \quad  3200 \\
    1250 - 1750 \quad  & \quad  0    
    \end{tabular}
    \caption{Time-varying power profile used in simulations
      reproducing the experiment of Oglesby~\cite{Oglesby2014}.}
    \label{tab:validationTest_Parameters}
\end{table}

\begin{table}[!ht]
  \centering
  \begin{tabular}{| c | c |}
    \hline
    \multicolumn{2}{|c|}{Simulation parameters} \\
    \hline
    Domain size (m$^3$)  & $ 0.1 \times 0.1 \times 0.025$ \\
    $T_0$ (K) & 297.5 \\
    $\omega_0$ (m) & $0.20$ \\
    \hline
  \end{tabular}
  \caption{Simulation parameters used to reproduce the experiment of Oglesby~\cite{Oglesby2014}.}
  \label{tab:validation-test-init-data}
\end{table}

\begin{figure}
    \centering
    \includegraphics[width=0.49\textwidth]{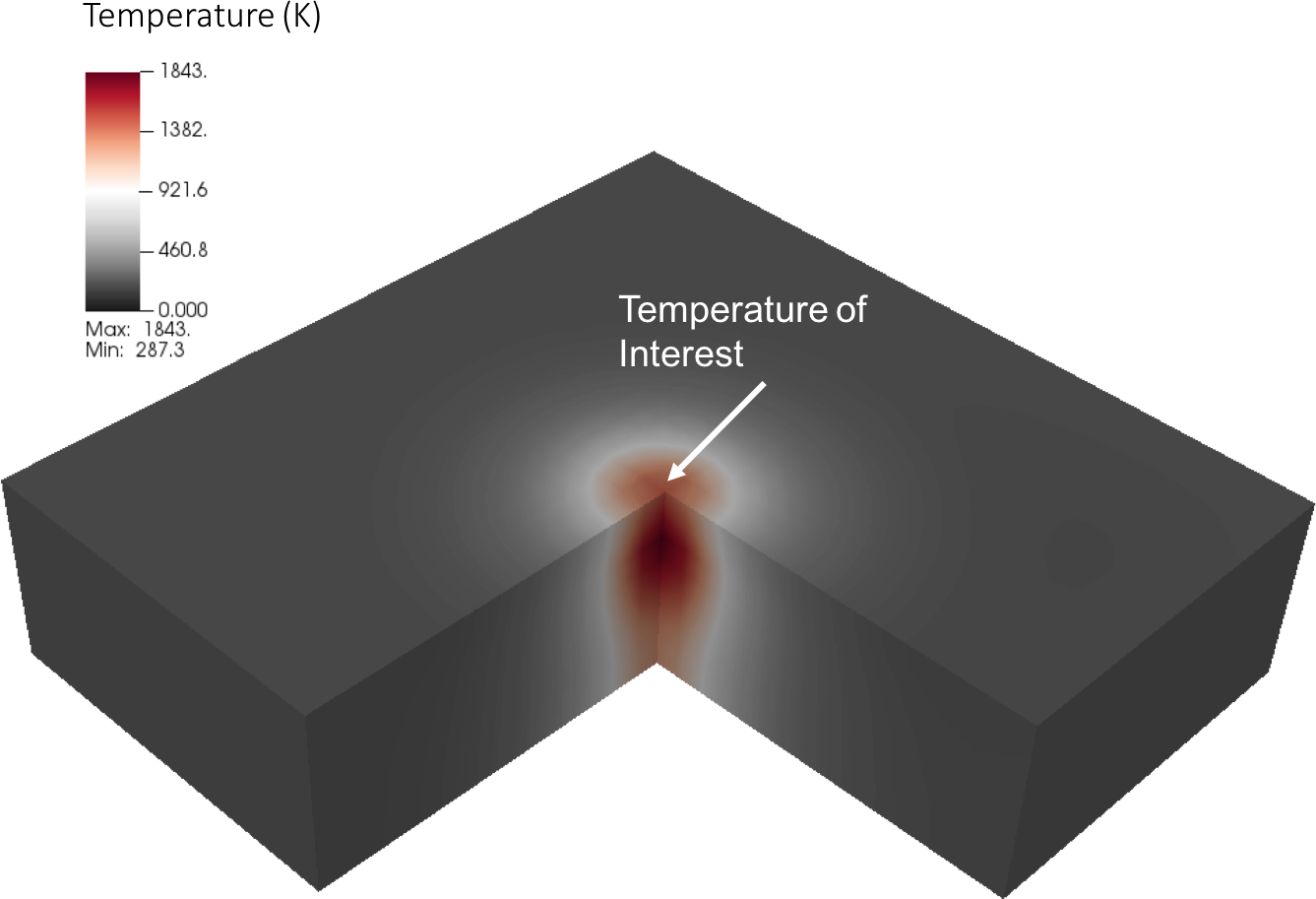}
    \caption{Illustration of the computational domain for comparison to
      the experiment of Oglesby~\cite{Oglesby2014}.  The domain is
      cuboidal and a section has been removed to illustrate the full
      temperature profile through the rock.  The arrow illustrates the
      centre of MMW beam where the surface temperature is recorded. }
    \label{fig:ValidationModelTemp}
\end{figure}

For this validation study, because the power was recorded at the
surface itself, the stand-off distance described in
section~\ref{sec:methodology} is set to zero.  In addition to
measuring the power incident on the surface, the temperature here was
also recorded at the centre of the incident source.  This was done by
measuring the thermal emission, $\varepsilon T$, using a 137 GHz
radiometer.  An illustration of the computational domain, and the
location at which temperature is computed in simulation results,
comparable to experiment, is given in
figure~\ref{fig:ValidationModelTemp}.

Material parameters are given by the relationships in
section~\ref{sec:methodology}.  Melt~\cite{Larsen} and
vaporisation~\cite{Bronshten1983} temperatures ($T_m$ and $T_v$
respectively) are given by
\begin{equation}
  \label{eq:granite-melt-vap}
  T_m = 1488 - 1533 \text{\ K}, \quad T_v = 3233 - 3503\text{\ K}.
\end{equation}
where the range in values results in differences in composition of the
rock; in this work the mid-point of these values is used.

Density of granite is given by equation~(\ref{eq:dens-eq}) where
\begin{equation}
  \label{eq:granite-den-par}
  a_\rho = -0.2877 \text{\ kg/(m$^3$K)}, \quad b_\rho = 2884 \text{\ kg/m$^3$}
\end{equation}
and above $T = 1273$ K, a constant value of $\rho = 2300$ kg/m$^3$ is
used.  Thermal conductivity is given by
equation~(\ref{eq:th-cond-rec}) where
\begin{equation}
  \label{eq:granite-cond-par}
  a_\kappa = 0.3154 \text{\ m K/W}, \quad b_\rho = 3.8 \times 10^{-4} \text{\ m/W}
\end{equation}
and above $T = 1073$ K, a constant extrapolation is used.  Specific heat
is given by equation~(\ref{eq:sp-heat}) with
\begin{equation*}
  a_c = -2.73 \times 10^4 \text{\ J K/kg}, \quad b_c = 0.894 \text{\ J/kg}
\end{equation*}
\begin{equation}
  \label{eq:granite-sp-param}
  c_c = 788 \text{\ J/(kg K)}, \quad d_c = 0.184 \text{\ J/kg}, \quad
  e_c = 1249 \text{\ J/(kg K)},
\end{equation}
with a cut-off temperature of $T_c = T_m$. Latent heat of fusion of
granite is $\xi_f = 3.4\times 10^5$ J/kg and since solidus and
liquidus temperatures are not available for granite, in order to
ensure these contributions occur over a sufficient temperature range
for algorithmic stability, latent heat is spread over 200 K around
melt temperature.  Latent heat of vaporisation is
$\xi_v = 4.8\times 10^6$ J/kg, and, as detailed in
section~\ref{sec:methodology}, it is spread over a range of
270 K from evaporation temperature onwards.  The absorptivity of solid
granite used in this work is $\alpha_\lambda = 14$ Np/m, and for
liquid granite, $\alpha_\lambda = 56$ Np/m.  Finally, the radiative
emissivity of granite is given by $\varepsilon = 0.7$ for $T \le T_m$
and $\varepsilon = 0.3$ for $T > T_m$, and a convective heat transfer
coefficient of $h = 56$ W/(m$^2$ K) is used, as given by Woskov and
Michael~\cite{Woskov2012} and Oglesby~\cite{Oglesby2014}.

\begin{figure}[h!]
  \centering
  \includegraphics[width=0.49\textwidth]{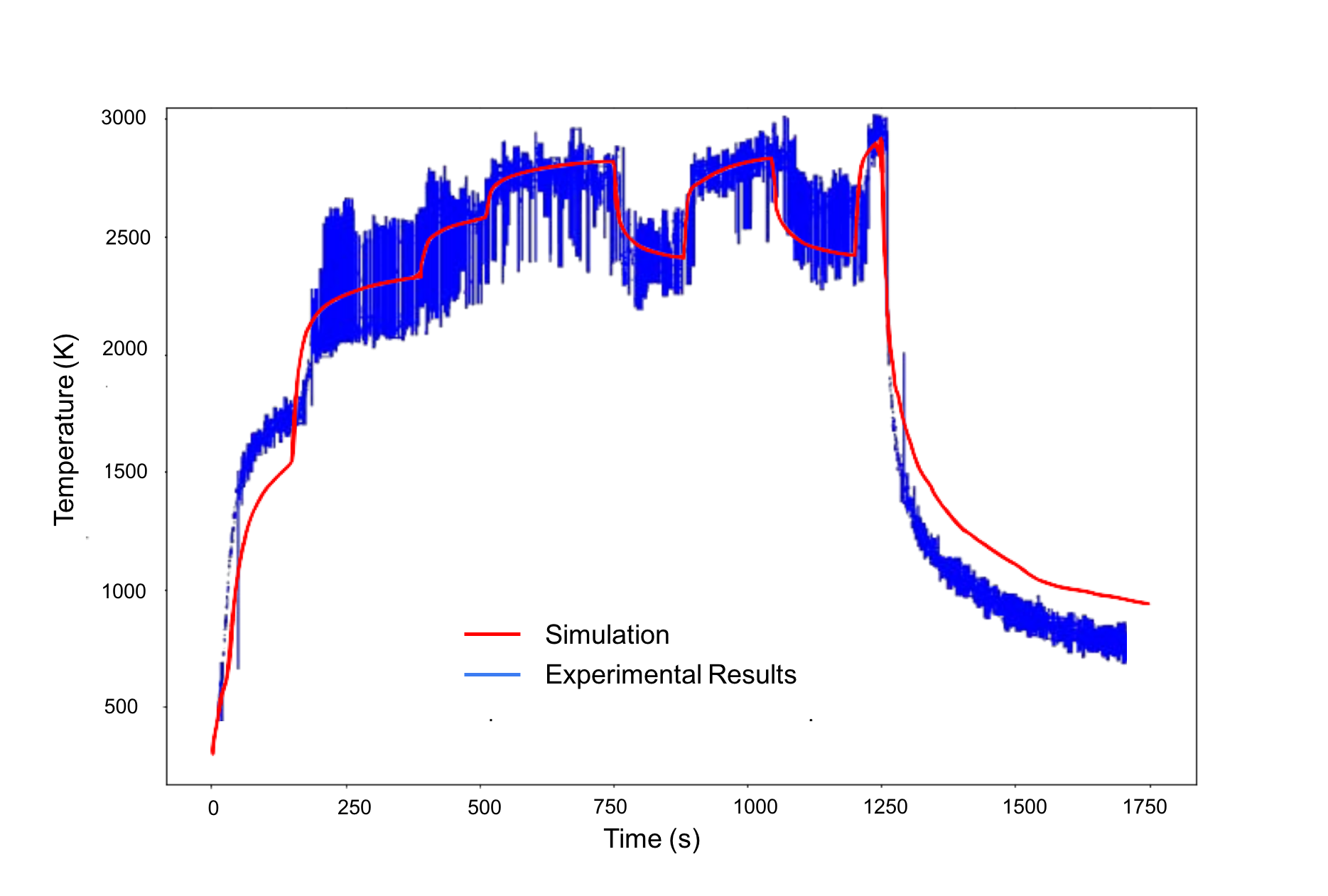}
  
  \caption{Comparison between experimental and simulation results for
    the temperature at the surface of the rock, directly beneath the
    MMW source.  It is noted that a correction to the experimental
    temperatures has been made to account for material
    emissivity~\cite{Oglesby2014}.  There is a good agreement between
    the two sets of results, the numerical model reproduced both the
    magnitude of the temperature, and the rise or fall behaviour as the
    incident power changes. }
  \label{fig:FullvalidationResults}
\end{figure}

The comparison between experimental and simulation results is shown in
figure~\ref{fig:FullvalidationResults}.  In order to make a direct
comparison between the two results, an emissivity correction is
required; the experimental measurements are recorded from the thermal
radiation above the surface, and hence the true temperature of the
surface is higher.  The true value of any such correction is not yet
known, and would require further experimental measurements, however, a
value of 0.7 is used in figure~\ref{fig:FullvalidationResults} and
this demonstrates a good agreement between simulation and experiment.
The results also demonstrate that the behaviour when the input power
is either increased or decreased matches the experiment.  When the
power changes, there is a sharp rise in temperature at the surface
which then tails off.  However, over this tailing off region, the
temperature continues to either rise, or fall, gradually (following
either an increase or a decrease respectively).  This behaviour is
captured in the simulation, suggesting the heating behaviour of the
material is correctly modelled.

\section{Evaluation of the model}
\label{sec:results}

\subsection{Single material configurations}
\label{sec:single-mater-results}

Having validated the numerical model, it is now used to evaluate the
material removal process for an isotropic, single material substrate.
In these test results, in order to ensure computational efficiency, a
very high source power is used.  Though unphysical for any current
gyrotron, the qualitative thermal behaviour will effectively be
unchanged from more realistic configurations, but will happen over
shorter time scales.  The simulation parameters are given in
table~\ref{tab:test-1-init-data}, with material properties as
described in section~\ref{sec:validation}.

\begin{table}[!ht]
  \centering
  \begin{tabular}{| c | c |}
    \hline
    \multicolumn{2}{|c|}{Simulation parameters} \\
    \hline
    Domain size (m$^3$)  & $1 \times 1 \times 2$ \\
    Simulation time (s) & $3.3$ \\
    $T_0$ (K) & 297.5 \\
    Material & Granite \\
    $P_0$ (W) & $1.7\times 10^{10}$ \\
    $\omega_0$ (m) & $0.22$\\
    \hline
  \end{tabular}
  \caption{Initial data for evaporation of granite under a
    high-powered MMW source.}
  \label{tab:test-1-init-data}
\end{table}

\begin{figure*}[btp]
    \centering
    \includegraphics[width=0.85\textwidth]{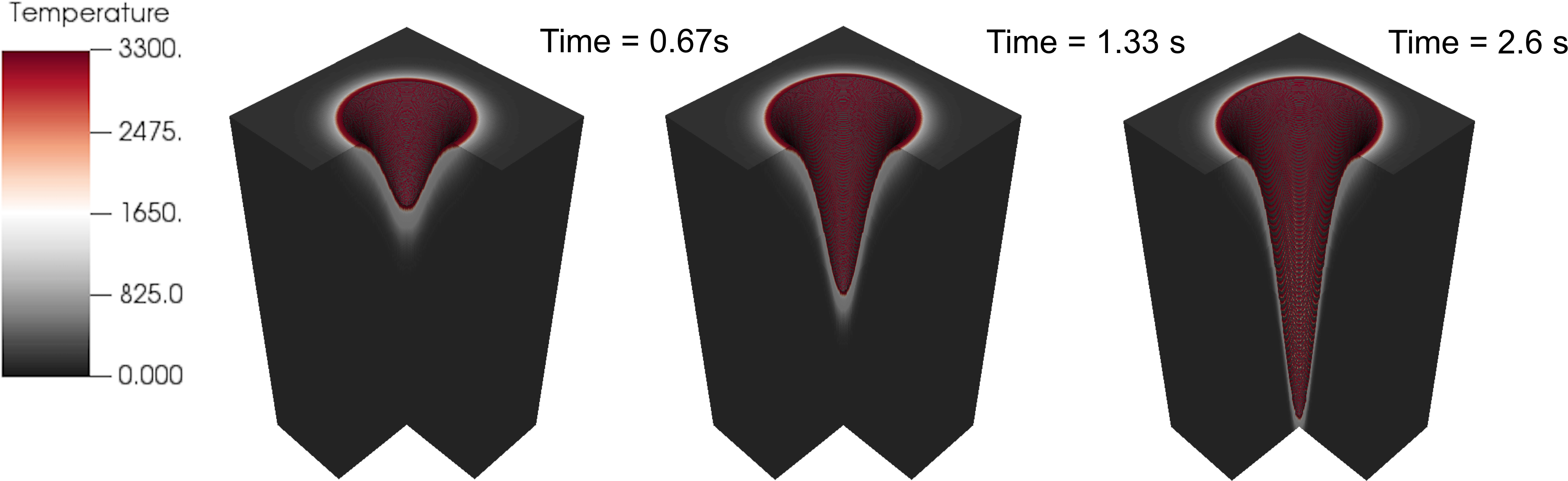}
    \caption{Temperature profile for granite under an MMW source with
      conditions as given in table~\ref{tab:test-1-init-data} over the
      course of the simulation.  A
      slice is removed from the domain to show internal behaviour.
      The high powered source removes material rapidly, and diffusion
      throughout the rock is minimal over these time scales.}
    \label{fig:MaterialRemova}
\end{figure*}

The temperature evolution within the granite for the test with initial
data described by table~\ref{tab:test-1-init-data} is shown in
figure~\ref{fig:MaterialRemova} at three snapshots in time.  In this
isotropic, single-material case, the geometry of the wellbore closely
correlates with the Gaussian shape of the MMW source, which has been
captured smoothly over the course of the simulation.  Due to the very
high power of the MMW source in this example, the material heating and
removal happens over a faster time scale than temperature diffusion,
hence there is a strong temperature gradient at the edges of the
wellbore, which is captured in a non-oscillatory manner by the
underlying numerical techniques.

\subsection{Effect of absorption coefficient on wellbore structure}
\label{sec:Rocks with different Absorption Coefficient }

The results shown in figure~\ref{fig:MaterialRemova} are for an
isotropic granite material, and show the absorptive properties of the
rock, as described in section~\ref{sec:validation}.  However, there
are many ways where absorptivity is altered, typically due to changes
in the emissivity or the electrical conductivity of the rock, which
govern the overall absorptivity, shown in
equation~(\ref{eq:absorptionCoefficientmethod}).  Additionally,
magnetic permeability may also have an effect, though this is only a
consideration in magnetised materials, which are less commonly
encountered.  When considering MMW drilling within granite, it is
common to encounter regions with high quartz content, which can be of
concern due to the low absorptivity of quartz.  Electrical
conductivity can be increased for a wider range of materials if the
rock is fractured and in particular if there is water saturation,
which would typically result in a rise in the absorption coefficient.

For geothermal drilling processes, the regions of low absorptivity are
of greatest concern; these could take longer to drill through, and
there may then be non-uniform heating with respect to the neighbouring
material, and thus a similarly non-uniform wellbore shape.  To
investigate the effects of this behaviour, a test simulation is
considered with a very low absorption coefficient,
$\alpha_\lambda = 0.7$ Np/m, and compared to the results shown in
figure \ref{fig:MaterialRemova}.

\begin{figure}
    \centering
    \includegraphics[width=0.49\textwidth]{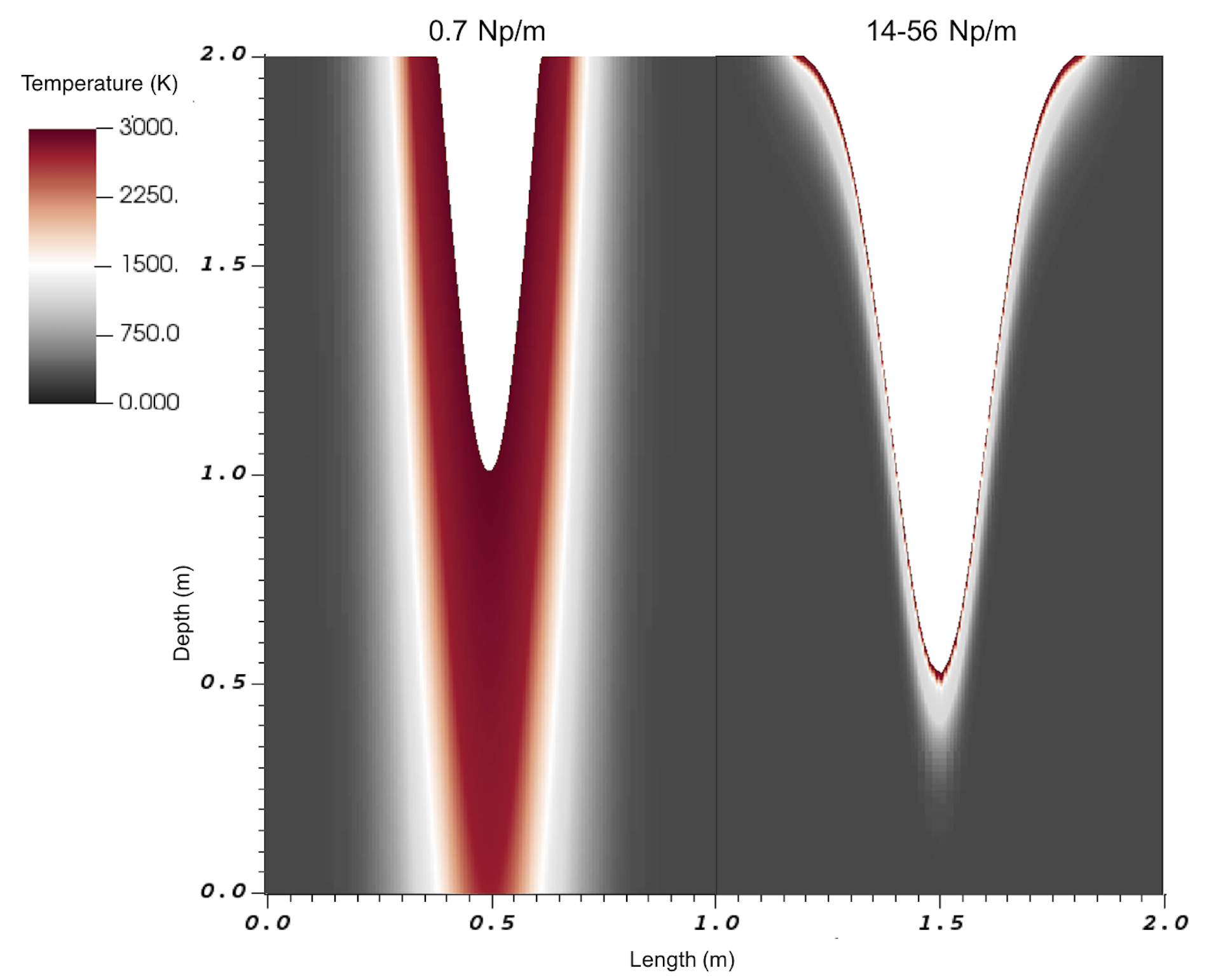}
    \caption{Comparison between the thermal profile of materials with
      different absorptivity, showing a slice through the centre of
      the computational domain.  Both results are shown after 3.3 s
      with the left plot showing a material with
      $\alpha_\lambda = 0.7$ Np/m, whilst the right plot shows granite
      values for $\alpha_\lambda$ given in
      section~\ref{sec:validation}.  The low absorptivity results in
      heating over a greater volume, and this subsequently results in
      a lower volume of material removed at a given time.}
    \label{fig:DownView}
\end{figure}

\begin{figure}
    \centering
    \includegraphics[width=0.49\textwidth]{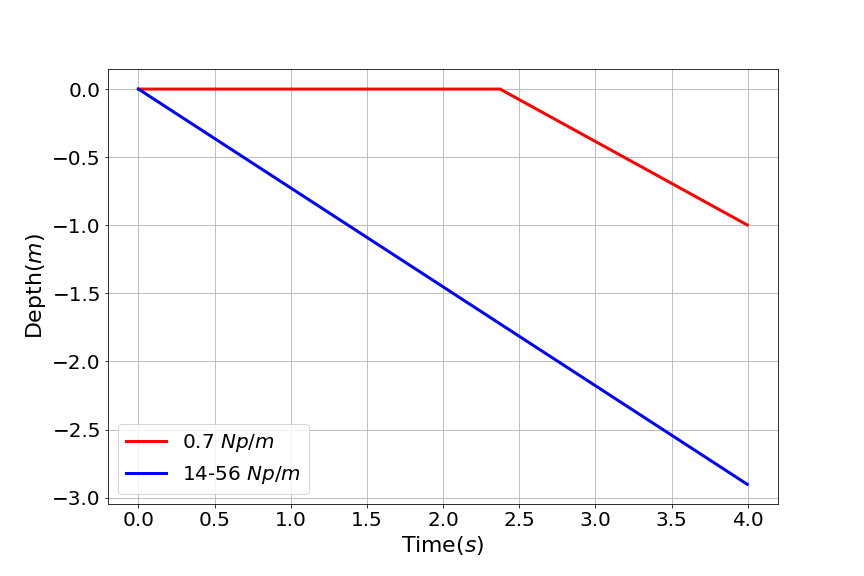}
    \caption{Penetration depth as a function of time for two different
      absorptivities.  Whilst the lower absorptivity material takes
      longer to start melting, the actual rate of penetration is then
      also slightly slower.}
    \label{fig:rate0fpeneratration}
\end{figure}

Figure~\ref{fig:DownView} shows the thermal profile for an
absorptivity of $\alpha_\lambda = 0.7$ Np/m compared to an equivalent
time for granite with the absorption properties given in
section~\ref{sec:validation}.  The high absorptivity shows a strong
thermal gradient at the surface, since power does not penetrate very
far into the material.  The lower absorptivity material has a much
longer penetration depth of the MMW source, and therefore a greater
volume of material is heated.  The temperature rise per unit volume is
then lower, and as a result, less material has been removed at a given
time.  Figure~\ref{fig:rate0fpeneratration} compares the rate of
penetration as a function of time.  Quantitatively, rate of
penetration, $\dot{d}$, is related to power density, $\varrho_p$ and
the specific energy of vaporisation, $e_v$, through
\begin{equation}
    \centering 
    \dot{d} = \frac{\varrho_p}{e_v} \quad \quad e_v = c_{p,m}  \left(T_v - T_m\right) + \xi_v
    \label{eq:ROP}
\end{equation}
where $c_{p,m}$ is the specific heat of melt~\cite{Tang2014}.  In
addition to the melting and vaporisation properties, this relationship
depends on absorptivity, entering though a reduction in power density
in the low-absorptivity case.  This is clearly apparent in
figure~\ref{fig:rate0fpeneratration}; once material reaches
evaporation temperature, the actual rate of removal is then slower for
the material with low absorptivity; for this high-power source, the
rate of penetration for granite is 0.73 m/s, whilst for the low
absorptivity material it is 0.61 m/s. For single, isotropic materials,
absorptivity is therefore not a significant concern beyond the
reduction in rate of penetration.  However, the behaviour for
anisotropic situations, where absorptivity is not constant in space,
may result in either insufficient or unwanted removal of material;
this can be considered through multi-strata simulations.

\section{Multi-strata modelling}
\label{sec:multi-strata-modell}

Whilst it is important to understand the evaporation behaviour, and to
predict the rate of penetration, through uniform rock, being able to
capture the behaviour at the transition between material or
configurations will be required to develop understanding of the MMW
drilling process.  The model described in
section~\ref{sec:methodology} is able to incorporate arbitrary numbers
and geometric configurations of materials, with material properties
required for solving equation~(\ref{eq:heat-eq}) defined through the
material scalar parameter.  To assess this capability, three tests are
considered, each with a planar material interface defined by the
surface
\begin{equation}
  \label{eq:planar}
  -0.5x + 0.2y-z + 1 = 0.
\end{equation}
The first test case considers a basalt-granite interface where basalt
is the upper material (the first material upon which MMWs are
incident), and all parameters for granite as given in
section~\ref{sec:validation}.

For basalt, melt~\cite{Larsen} and vaporisation~\cite{Bronshten1983} temperatures are given by
\begin{equation}
  \label{eq:basalt-melt-vap}
  T_m = 1257 - 1533 \text{\ K}, \quad T_v = 3233 - 3503\text{\ K}.
\end{equation}
where, as with granite, the range is dependent on the actual
composition of the basalt, hence again the mid-point value is used.
The density follows equation~(\ref{eq:dens-eq}) with constants given
by
\begin{equation}
  \label{eq:basalt-den-param}
  a_\rho = -0.06 \text{\ kg/(m$^3$K)}, \quad b_\rho = 2897 \text{\ kg/m$^3$}
\end{equation}
and above $T = 933$ K, a constant value of $\rho = 2801$ kg/m$^3$ is
used.  Thermal conductivity is given by
equation~(\ref{eq:th-cond-eq}) where
\begin{equation*}
  c_\kappa = 1.719\times 10^{-6} \text{\ W K/m}, \quad d_\rho = -3.588 \times 10^{-4} \text{\ W/m}
\end{equation*}
\begin{equation}
  \label{eq:basalt-therm-param}
  e_\rho = 3.071 \text{\ W/(m K)}
\end{equation}
and above $T = T_m$, a constant extrapolation is used.  Specific heat
follows equation~(\ref{eq:sp-heat}) with
\begin{equation*}
  a_c = -7.52 \times 10^4 \text{\ J K/kg}, \quad b_c = 1.43 \text{\ J/kg}
\end{equation*}
\begin{equation}
  \label{eq:basalt-sp-param}
  c_c = 415 \text{\ J/(kg K)}, \quad d_c = 0.176 \text{\ J/kg}, \quad
  e_c = 924 \text{\ J/(kg K)},
\end{equation}
with a cut-off temperature of $T_c = T_m$.  Latent heat of fusion of
basalt is $\xi_f = 4.2\times 10^5$ J/kg and, as with granite, this is
spread over 200 K around melt temperature, and latent heat of
vaporisation is $\xi_v = 4.0\times 10^6$.  The absorptivity of solid
basalt is $\alpha_\lambda = 14$ Np/m, and for
liquid basalt, $\alpha_\lambda = 28$ Np/m.  The parameters for
radiative and convective losses in basalt have not been measured
extensively, hence in this work, we assume that  $\varepsilon = 0.5$
and that the convective heat loss coefficient of granite, $h = 56$
W/(m$^2$ K), can be used for basalt.

This configuration is motivated by the fact that basalt formed from
volcanic eruptions overlays granite which originated from magma
intrusion.  The second test uses an absorptivity of
$\alpha_\lambda = 0.7$ Np/m for basalt, investigating the effects of a
low-absorptivity region on top of a high-absorptivity one.  The third
test reverses this configuration, using $\alpha_\lambda = 0.7$ Np/m
for granite instead.  The remaining initial data for these
configurations is given in table~\ref{tab:test-2-init-data}.

\begin{table}[!ht]
  \centering
  \begin{tabular}{| c | c |}
    \hline
    \multicolumn{2}{|c|}{Simulation parameters} \\
    \hline
    Domain size (m$^3$)  & $1 \times 1 \times 1$ \\
    Simulation time (s) & $3$ \\
    $T_0$ (K) & 297.5 \\
    $P_0$ (W) & $10^{9}$ \\
    $\omega_0$ (m) & $0.22$\\
    \hline
  \end{tabular}
  \caption{Initial data for evaporation of a multi-strata configuration.}
  \label{tab:test-2-init-data}
\end{table}

\begin{figure*}[btp]
    \centering
    \includegraphics[width =0.75\textwidth]{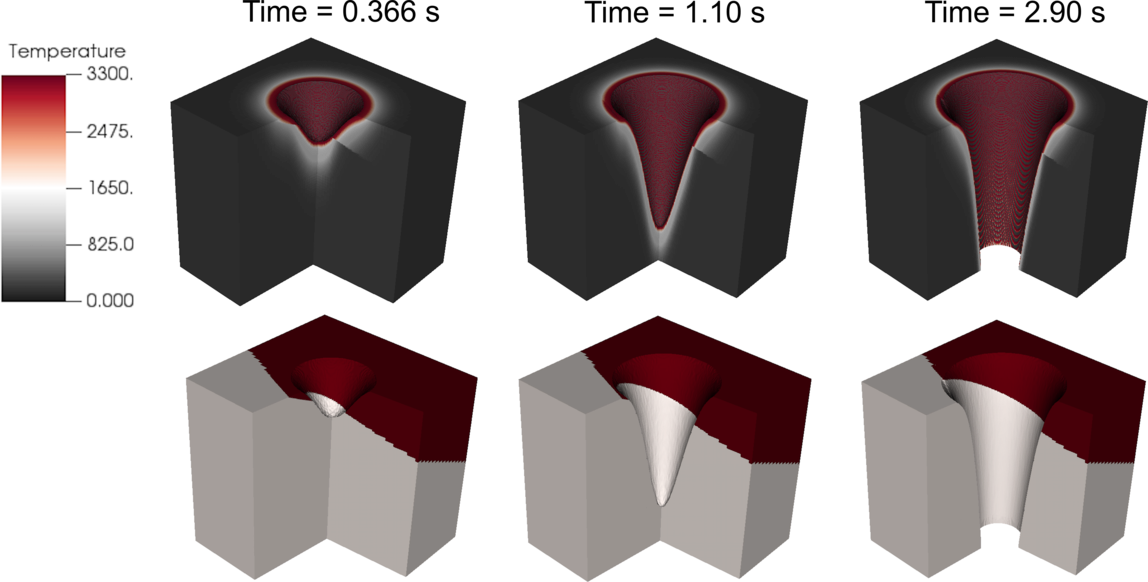}
    \caption{Simulation results for a basalt layer on top of
      granite. A section of the domain is removed to show internal
      behaviour of temperature (top), and the material, were red shows
      basalt and beige shows granite (bottom). The wellbore follows a
      Gaussian profile with minimal change at rock interface. However,
      the effects of the change in material are visible in
      temperature, with a discontinuity along basalt-granite
      interface, which correlates with an increase in MMW absorptivity
      moving from molten basalt to granite. }
    \label{fig:BasaltonGranite}
\end{figure*}

\begin{figure}
    \centering
    \includegraphics[width=0.49\textwidth]{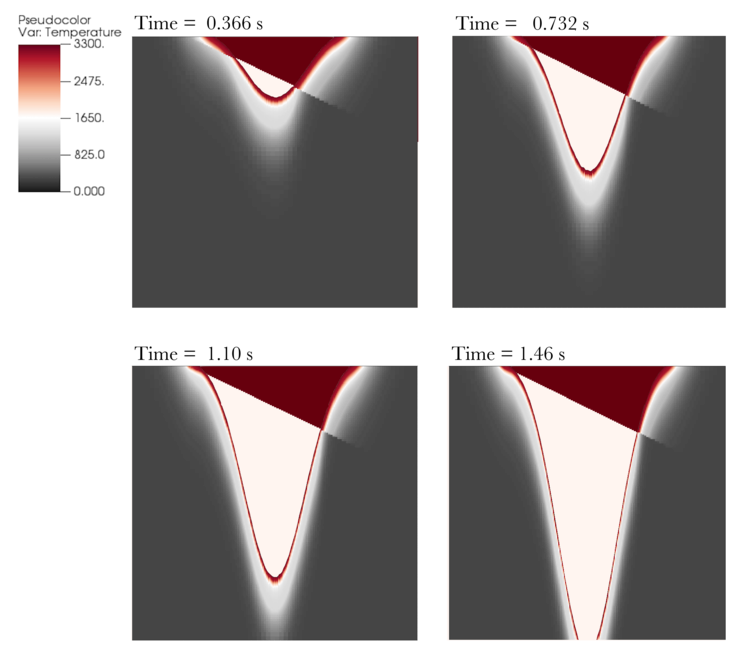}
    \caption{Cross section through the centre of the wellbore for the
      case of a basalt layer on granite, shown in
      figure~\ref{fig:BasaltonGranite}.  The temperature profile is
      shown for the solid and liquid phases, whilst the material (and
      hence interface) is plotted within the wellbore itself.  The
      change in temperature gradient between the two materials is
      clear, though the shape of the wellbore remains largely
      symmetric.}
    \label{fig:BasaltonGraniteCrossSection}
\end{figure}

Figure \ref{fig:BasaltonGranite} shows the simulation for basalt on
granite.  The two materials have comparable properties, and this shows
in the relatively smooth nature of the wellbore.  However, the lower
absorptivity of molten basalt is apparent in the slightly wider
high-temperature region, i.e.\ there is a shallower temperature
gradient.  This behaviour is visible more clearly in
figure~\ref{fig:BasaltonGraniteCrossSection}, showing a cross section
through the centre of the domain.  These results, along with the
implications for the rate of penetration from
figure~\ref{fig:rate0fpeneratration}, suggest that these interfaces
between similar igneous rock types will not be a major issue for MMW
drilling applications.

\begin{figure*}
    \centering
      \includegraphics[width=0.6\textwidth]{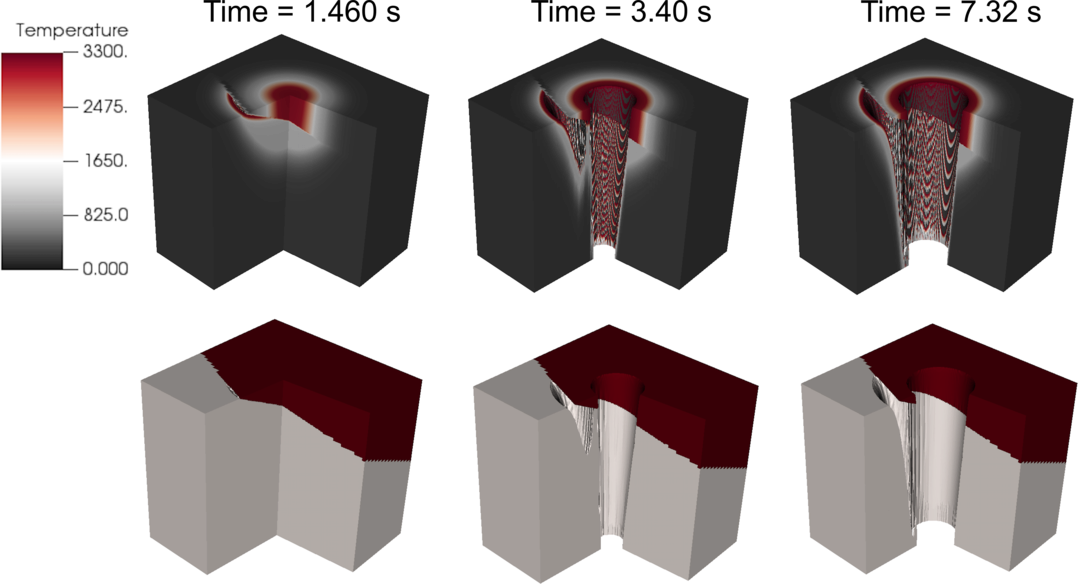}
     \includegraphics[width=0.6\textwidth]{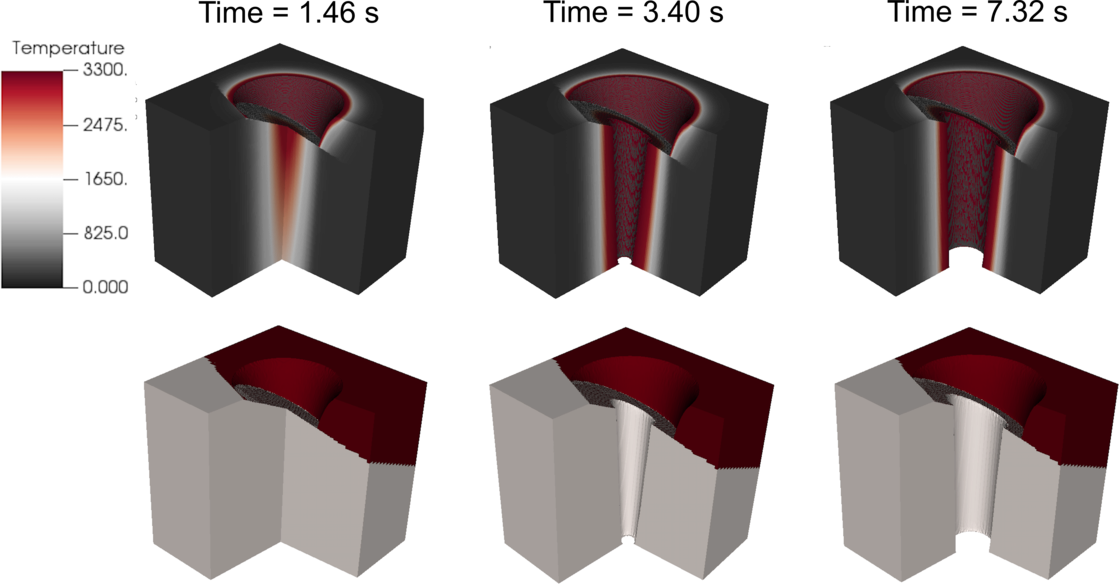}
     \caption{ Simulation results for the multi-strata test cases with
       a low-absorptivity region; the top figure shows this region on
       top of granite, whilst the bottom figure shows basalt on top of
       a low absorptivity layer.  In the top figure, it is clear that
       the radial growth of the wellbore is inhibited by the
       low-absorptivity layer.  In the bottom case, the lower rate of
       expansion leads to the continued widening of the borehole
       through the basalt region on top. }
    \label{fig:Multistrataoverall}
\end{figure*}

In figure~\ref{fig:Multistrataoverall}, the effects of replacing one
of the strata with a low absorptivity material are demonstrated.  In
the top image in this figure, a case of a low absorptivity region
above granite is shown.  In this case, there is an overall reduction
in the wellbore diameter and this corresponds to a slower evaporation
of the granite beneath this layer.  In general, the energy absorption
by the low absorptivity layer is sufficient to prevent any direct melt
of the granite beneath it.  However, where this layer is sufficiently
thin, then melting initially occurs beneath the surface.  The model
presented within this work can predict the onset of such behaviour,
though at this point, additional techniques may be required to
correctly simulate the resultant behaviour.  In particular, a pressure
build-up would be expected within this region, which is not accounted
for in this thermal approach.

The third test case, in which the lower material has low absorptivity,
is shown in the bottom image in figure~\ref{fig:Multistrataoverall}.
In this case, at early times, the expected behaviour is observed, in
which the basalt region is removed much more rapidly, and then a new,
angled surface heats up. At later times, the heat flux from the MMW
source mostly interacts with the low absorptivity surface, though
there is continued widening of the wellbore in the basalt region.  In
this configuration, the radius of the wellbore in this basalt region
would then be larger than expected, which may have implications on
equipment placed down the borehole.

These two test cases, though considering idealised materials,
demonstrate that the model presented here can deal with sudden changes
in material behaviour.  In particular, it is possible to predict the
effects of a change in one material on both the local rate of
penetration behaviour, but also in the effects on neighbouring
materials.  In all cases, the numerical model has demonstrated that it
can robustly deal with material changes within the domain.  These
changes are handled in a smooth manner, without generating
oscillations.  This gives confidence that this model can now be
applied for geologically realistic configurations.

\section{Conclusions}
\label{sec:conclusions}

In this work, a thermal model for material evaporation through an MMW
source has been presented, with a view to an application for drilling
through geological materials.  Of particular interest is igneous rock,
for which conventional drilling methods struggle, and this has formed
the basis of the validation, and subsequent evaluation of the model.

Validation was based on the experiment of Oglesby~\cite{Oglesby2014},
for which a slab of granite was subjected to heating through a
kilowatt MMW gyrotron source.  This heating lasted 30 minutes, and the
incident power was changes several times over the course of the
experiment.  It was demonstrated in section~\ref{sec:validation} that
the model presented here could match the thermal profile recorded from
this experiment; both temperature values and the equilibration of
these values as surface heat losses matched the incident power were
shown.

The model was then utilised in section~\ref{sec:single-mater-results}
to investigate effects of evaporation of materials with differing
absorptive properties.  It was identified that a lower absorption
coefficient delayed the onset of evaporation, and subsequently
resulted in a slower rate of penetration.  One of the key advantages
of this model is that it allows for an arbitrary geometric
configuration of different rock types, or regions with different
physical properties.  This was investigated in
section~\ref{sec:multi-strata-modell}, where a typical
basalt-on-granite configuration was studied, as well as a more extreme
case, with a strata of low-absorptivity rock.  This latter case is of
interest for cases such as granite with high quartz content; quartz
allowing for more transmission of MMW radiation.  In this case, the
multi-strata simulations were handled smoothly, and able to
demonstrate the differences in wellbore structure in all cases.

The work presented here highlights several areas for further study, as
well as additional developments to further increase the applicability
of the model.  Two types of rock, basalt and granite, were considered
here, and success at the model for dealing with these materials means
work can in future be extended to more rock types.  However, this
extension, as well as improving accuracy of the current results,
requires further experimental study into the material properties of
rocks at high temperatures.  As discussed in
section~\ref{sec:methodology}, the properties of these materials are
generally only well-studied up to around 1000 K, whilst the MMW
drilling approach is expected to evaporate the rock, with temperatures
exceeding 3000 K.  Simple constant or linear extrapolations of the
known data were used in this work, and were demonstrated to produce a
good physical description of the material.  However, extending the
knowledge of the thermal properties of rocks to higher temperature
ranges, and to more rock types, would have an obvious benefit to the
results presented in this work.

In section~\ref{sec:surface-heat-losses}, the importance of the
surface treatment was considered.  Within the thermal approach, the
heating and evolution of gas needs to be imposed as a surface
condition, a variety of options were considered, guided by the
experimental results of Oglesby~\cite{Oglesby2014}.  These conditions
could, in future, be augmented by coupling the model to a gas flow
simulation, which would allow the effect on surface cooling of imposed
gas flow scenarios to be simulated.  This could then identify optimal
conditions for rate of penetration whilst preventing plasma formation
in the gas beneath the wave guide.

The work presented here has assumed that the absorption coefficient,
$\alpha_\lambda$ is constant, or at least piece-wise constant.  This
is an assumption that simplifies the power input; allowing it to be
expressed as a function of height only.  In practice, the absorption
coefficient depends on temperature, and as a result there is a
non-uniform absorption of power with depth.  This requires an integral
approach, where power absorbed is not dependent only on the power
incident on the surface, but of that absorbed by material above the
height of computation.  This can be challenging, given that the height
of the surface is not constant; the power absorbed needs to be
integrated with respect to depth for every point in the
$(x,y)$-plane.  Within this framework, such a technique could be
incorporated into future work, taking advantage of the fast-sweeping
algorithm for the signed distance function; this algorithm itself
performs an extrapolation from the surface outwards, and hence the
computational framework exists to achieve this extension.  This has
the additional advantage that power absorption can be computed even
for complex AMR grid structures, where information transfer concerns
exist.

This proposed future development would also allow inhomogeneities
within the rock strata to be considered.  Here, local changes in
properties, including absorption, could be incorporated, either as
distinct crystalline structures, or a macroscopic averaging, depending
on the length-scales required.  The generation of suitable anisotropic
structures has been considered at crystal grain-scales by Toifl et
al.~\cite{Toifl2016} and Quey et al.~\cite{Quey2011}, whilst larger
scale approximations have been considered by Tang et
al.~\cite{Tang2014}  In addition to the development proposed for
absorptive properties, the scalar parameter for modelling material
type used in this work would allow these techniques to be incorporated
into the model presented here.

These considerations for future work would enhance the capabilities of
this work to simulate the evaporation complex geological materials
under a MMW source.  The work described within this paper describes a
novel technique which underlies these ideas, and itself augments the
experimental studies of MMW drilling using gyrotrons, allowing for a
rapid investigation of parameter space, both for the source and the
substrate.  The adaptive meshing approach, which does not depend on
assumptions as to the wellbore shape or growth rate, allows for
computationally efficient, large-scale simulations.

\begin{acknowledgments}
  The authors would like to thank Franck Monmont of Quaise Energy for
  technical input throughout this work.
\end{acknowledgments}





\bibliography{refs}

\end{document}